\documentclass[9pt,twocolumn,twoside]{opticajnl}
\journal{opticajournal} 
\pdfoutput=1

\setboolean{shortarticle}{true}

\usepackage{braket}
\usepackage{lineno}
\usepackage{times}
\usepackage{amsmath}
\usepackage{indentfirst}
\usepackage{float}
\usepackage{multirow}
\usepackage{makecell}
\usepackage{graphicx}
\usepackage{amsthm,amssymb,bbm,newtxmath}
\usepackage{mathrsfs}
\usepackage[noend]{algpseudocode}
\usepackage{array}
\usepackage{booktabs}
\usepackage{stfloats}
\usepackage{makecell}
\usepackage{lineno}

\title{Quantum generative adversarial learning in photonics}

\author[1,$\dagger$]{Yizhi Wang}
\author[1,$\dagger$]{Shichuan Xue}
\author[1]{Yaxuan Wang}
\author[1]{Yong Liu}
\author[1]{Jiangfang Ding}
\author[1]{Weixu Shi}
\author[1]{Dongyang Wang}
\author[1]{Yingwen Liu}
\author[1]{Xiang Fu}
\author[1]{Guangyao Huang}
\author[1]{Anqi Huang}
\author[1]{Mingtang Deng}
\author[1,*]{Junjie Wu}

\affil[1]{Institute for Quantum Information \& State Key Laboratory of High Performance Computing, College of Computer Science and Technology, National University of Defense Technology, Changsha 410073, China}

\affil[$\dagger$]{These authors contributed equally to this work.}
\affil[*]{junjiewu@nudt.edu.cn}

\begin{abstract}
Quantum Generative Adversarial Networks (QGANs), an intersection of quantum computing and machine learning, have attracted widespread attention due to their potential advantages over classical analogs. However, in the current era of Noisy Intermediate-Scale Quantum (NISQ) computing, it is essential to investigate whether QGANs can perform learning tasks on near-term quantum devices usually affected by noise and even defects. In this Letter, using a programmable silicon quantum photonic chip, we experimentally demonstrate the QGAN model in photonics for the first time, and investigate the effects of noise and defects on its performance. Our results show that QGANs can generate high-quality quantum data with a fidelity higher than 90\%, even under conditions where up to half of the generator's phase shifters are damaged, or all of the generator and discriminator's phase shifters are subjected to phase noise up to 0.04$\pi$. Our work sheds light on the feasibility of implementing QGANs on NISQ-era quantum hardware.
\end{abstract}

\setboolean{displaycopyright}{true} 

\begin{document}

\maketitle
\textit{Introduction}.—
The generation of quantum states plays a fundamental role in quantum computing \cite{Nielsen2010}.
Using the power of machine learning techniques \cite{Cerezo2022}, quantum generative adversarial networks (QGANs) can generate high-quality quantum states without requiring knowledge of the explicit structure of the states \cite{Lloyd2018a,Dallaire-Demers2018}.
QGANs combine the power of quantum mechanics with the ingenuity of generative adversarial networks \cite{Goodfellow2014}, and are characterized by their potential for generating quantum data in a wide range of fields, from  real-world classical distribution loading \cite{Zoufal2019,Situ2020,Romero2021} and images generation \cite{Huang2021a} to drug discovery \cite{Li2021a}. 
In terms of experiments, several QGANs capable of generating high-quality quantum states have been demonstrated in superconducting circuits \cite{Hu2019,Huang2021,Huang2021a,Ahmed2021,Anand2021a,Niu2022b} and trapped-ion processors \cite{Zhu2022}. 
These previous results are impressive; however, whether QGANs can work normally on NISQ-era devices, which are ubiquitously plagued by noise and defects \cite{Preskill2018,Arute2019,Gong2021,Wang2018,Qiang2021}, is in urgent need of further investigation. 
Especially in photonics, the research on the performance of QGANs on photonic devices is still lacking.

In this Letter, we have experimentally implemented QGANs in photonics for the first time. Using a programmable silicon quantum photonic chip, we have implemented QGANs that the generator and discriminator are both programmable quantum circuits. Through experimental demonstrations and numerical simulations, we study several learning processes with QGANs that are subjected to different levels of noise. We also investigate the robustness of QGANs in defective chips.
Our results demonstrate the reliability of QGANs on non-fault-tolerant devices and open up a vista for practical applications of QGANs with photons in the NISQ era.

\begin{figure*}[t]
  \centering\includegraphics{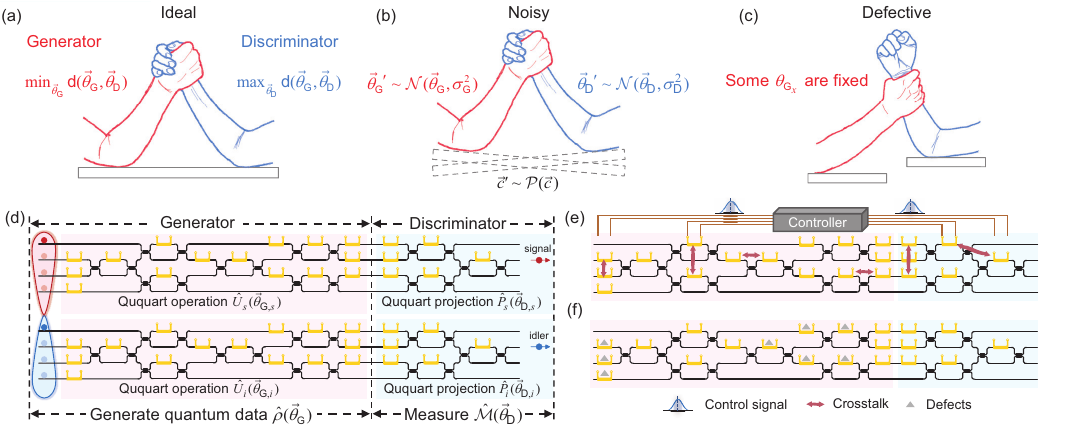}
  \caption{(a)-(c) A comparison of the arm-wrestling in QGAN training on ideal, noisy, and defective quantum devices. (a) Ideally, the competition should take place on a sturdy table, where the Generator tries to minimize $\text{d}(\vec{\theta}_{\text{G}},\vec{\theta}_{\text{D}})$ while the Discriminator attempts to maximize $\text{d}(\vec{\theta}_{\text{G}},\vec{\theta}_{\text{D}})$. (b) The noise causes movement in the table, which is disruptive to the competition. (c) The generator's defects place it at a disadvantageous position in the competition.
  (d)-(f) Schematic of our quantum photonic chip and the phase shifters' noise and defects. 
  (d) We use two universal linear optical circuits that perform single-ququart operations $\hat{U}_s(\vec{\theta}_{{\text{G}},s})$ and $\hat{U}_i(\vec{\theta}_{{\text{G}},i})$ on the two maximally-entangled ququarts to act generator's circuit and generate the candidate state $\hat{\rho}(\vec{\theta}_{\text{G}})$. The following triangular circuits that perform single-ququart projections $\hat{P}_s(\vec{\theta}_{{\text{D}},s})$ and $\hat{P}_i(\vec{\theta}_{{\text{D}},i})$ are employed to implement discriminator's measurements $\mathcal{\hat{M}}(\vec{\theta}_{\text{D}})$.
  (e) The thermo-optic phase shifters are subjected to errors of two sources: electronic control deviation
  and residual crosstalk from other phase shifters. These errors result in the Gaussian noise in the phases $\vec{\theta}^{\prime} \sim \mathcal{N}(\vec{\theta},\sigma^2)$. 
  (f) The defects in generator's circuits. We will lose control of the damaged phase shifters, of which the phases are generally fixed values.  
  }
  \label{fig: setup}
\end{figure*}

\textit{The QGAN model subjected to noise and defects}.—The core idea behind our QGAN is a two-player adversarial game comprising a \emph{Generator} and a \emph{Discriminator}. Each of the two players has a variational quantum circuit: the generator circuit with parameters $\vec{\theta}_{\text{G}}$ and the discriminator circuit with parameters $\vec{\theta}_{\text{D}}$. 
At first, a true data is given, which is described by a density operator $\hat{\tau}$. The generator uses its circuit to produce a candidate quantum state $\hat{\rho}(\vec{\theta}_{\text{G}})$ that aims to imitate the true state $\hat{\tau}$ and fool the discriminator. The discriminator performs measurements $\mathcal{\hat{M}}(\vec{\theta}_{\text{D}})$ on its input state. Based on the measurement result, the discriminator determines whether the generated state is the true state $\hat{\tau}$ or not.

As illustrated in Fig. \ref{fig: setup}(a), we can make an analogy between QGAN training and arm wrestling, where the generator and discriminator improve themselves by competing against each other on the difference in the measurement results 
\begin{equation}
  \text{d}(\vec{\theta}_{\text{G}},\vec{\theta}_{\text{D}})=\left|\text{tr}[\mathcal{\hat{M}}(\vec{\theta}_{\text{D}})\hat{\rho}(\vec{\theta}_{\text{G}})]-\text{tr}[\mathcal{\hat{M}}(\vec{\theta}_{\text{D}})\hat{\tau}]\right|.
\end{equation}
In each round, the discriminator and generator take turns in exerting their strength.
The discriminator first unleashes his power by optimizing its parameters $\vec{\theta}_{\text{D}}$ in order to maximize $\text{d}(\vec{\theta}_{\text{G}},\vec{\theta}_{\text{D}})$, making it easier to distinguish between $\hat{\tau}$ and $\hat{\rho}(\vec{\theta}_{\text{G}})$ from the measurements.
Next, the generator's turn comes, and it updates the parameters $\vec{\theta}_{\text{G}}$ to prepare a new synthetic state $\hat{\rho}(\vec{\theta}_{\text{G}})$ that can minimize $\text{d}(\vec{\theta}_{\text{G}},\vec{\theta}_{\text{D}})$. A perfect evolution of the generator results in $\text{d}(\vec{\theta}_{\text{G}},\vec{\theta}_{\text{D}}) \approx 0$, which makes the generator more adept at replicating $\hat{\tau}$. 
By iterative competitions, the QGAN solves the minimax problem \cite{Lloyd2018a,Dallaire-Demers2018}
\begin{equation}
  \min\nolimits_{\vec{\theta}_{\text{G}}}\max\nolimits_{\vec{\theta}_{\text{D}}}\text{d}(\vec{\theta}_{\text{G}},\vec{\theta}_{\text{D}}).
\end{equation}
Ideally, the competition can eventually approach the unique Nash equilibrium \cite{Nashequilibrium,Lloyd2018a,Dallaire-Demers2018}, where the generator counterfeits the true state $\hat{\tau}$ to ensure $\text{d}(\vec{\theta}_{\text{G}},\vec{\theta}_{\text{D}}) \approx 0$ for arbitrary measurements $\mathcal{\hat{M}}(\vec{\theta}_{\text{D}})$, and the discriminator is unable to tell the measurement difference between the true state $\hat{\tau}$ and the generated state $\hat{\rho}(\vec{\theta}_{\text{G}})$.
It suggests that the generator has the ability to reproduce the true data $\hat{\tau}$ correctly without
performing quantum state tomography.

However, in the NISQ era, ubiquitous noise will have a direct impact on this competition. 
Noise can affect the actual hardware parameters and then the measurement results. As shown in Fig. \ref{fig: setup}(b), just like arm-wrestling on a wobbling table, dynamic noise can cause instability during the training of QGANs \cite{Borras2023}.
On photonic chips such as in Fig. \ref{fig: setup}(d), although photons are well known for not having problematic interactions with the environment that would lead to decoherence, phase noise caused by electronic control deviation and residual crosstalk from thermo-optic phase shifters are the primary sources of noise \cite{Paesani2017a,Qiang2021}, which is illustrated schematically in Fig. \ref{fig: setup}(e). 
However, it is hard to analytically characterize all the factors that cause phase noise on current photonic chips with high integration density, large number of components, and complex structures. Instead, the errors caused by these factors and even other unknown factors can be described by replacing the correct phases $\vec{\theta}$ with a random vector $\vec{\theta}^{\prime}$ sampled from a Gaussian distribution $\vec{\theta}^{\prime} \sim \mathcal{N}(\vec{\theta},\sigma^2)$ \cite{Paesani2017a}. 
In addition, due to the low rate of multi-photon generation \cite{Adcock2021}, shot noise in photon detection is another type of noise that can degrade the measurement accuracy and disturb the learning process. Shot noise can be described by replacing the actual number of photons $\vec{c}$ with a random vector $\vec{c}^\prime$ sampled from a Poissonian distribution $\vec{c}^\prime \sim \mathcal{P}(\vec{c})$ \cite{Qiang2021}.

Besides, it is common for near-term quantum devices to have defects, even in state-of-the-art superconducting processors \cite{Arute2019,Gong2021} and quantum photonic chips \cite{Wang2018,Qiang2021}. 
As illustrated in Fig. \ref{fig: setup}(c), the defects in the generator's circuit put it at a disadvantageous position in the competition, particularly weakening its ability to generate quantum data.
In photonic circuits shown in Fig. \ref{fig: setup}(f), the most severe defects are the broken phase shifters caused by electrical damage. 
These broken phase shifters are not adjustable and their phases are fixed.

\textit{Experimental methods}.—
We use a programmable silicon quantum photonic chip \cite{Xue2022,Wang2023} to implement the QGAN model and investigate QGAN's robustness against noise and defects. 
The schematic of the chip is presented in Fig. \ref{fig: setup}(d), which can prepare arbitrary two-ququart maximally entangled states and perform two-ququart projective measurements. 
Firstly, on-chip signal-idler photon pairs are produced \cite{Xue2022,Silverstone2014}.
Post-selecting the cases in which the signal photons exit at the top circuit and the idler photons exit at the bottom yields the two-photon maximally entangled state $\ket{\psi_0}=\frac{1}{2}\sum_{k=1}^4 {\ket{k}_s\otimes\ket{k}_i}$. Here, $\ket{k}_s$ and $\ket{k}_i$ represent the path modes of the signal and idler photons.
Then, the signal and idler photons are sent to universal linear optical circuits (background colored in pink) formed by Mach-Zehnder interferometers (MZIs) and phase shifters in the rectangular configuration \cite{Clements2016}. We can perform arbitrary single-ququart unitary transformations, $\hat{U}_s(\vec{\theta}_{{\text{G}},s})$ and $\hat{U}_i(\vec{\theta}_{{\text{G}},i})$, on each of the two photons. Then the evolved two-photon state becomes
\begin{equation} \label{Eq: Schmidt decomposition}
  \ket{g(\vec{\theta}_{\text{G}})}=\frac{1}{2} \hat{U}_s(\vec{\theta}_{{\text{G}},s}) \otimes \hat{U}_i(\vec{\theta}_{{\text{G}},i})\sum\nolimits_{k=1}^4 {\ket{k}_s\otimes\ket{k}_i}.
\end{equation}
According to the \emph{Schmidt decomposition} \cite{Nielsen2010}, Eq. (\ref{Eq: Schmidt decomposition}) can describe arbitrary two-ququart maximally entangled states. Therefore, our chip can prepare an arbitrary two-ququart maximally entangled state by configuring $\hat{U}_s(\vec{\theta}_{{\text{G}},s})$ and $\hat{U}_i(\vec{\theta}_{{\text{G}},i})$.
Together with the following measurement circuits formed by two triangular networks of MZIs, a projective measurement can be described by the two-ququart projector
\begin{equation}
  \mathcal{\hat{M}}(\vec{\theta}_{\text{D}})=\hat{P}_s(\vec{\theta}_{{\text{D}},s}) \otimes \hat{P}_i(\vec{\theta}_{{\text{D}},i}),
\end{equation}
where $\hat{P}_s(\vec{\theta}_{{\text{D}},s})$ and $\hat{P}_i(\vec{\theta}_{{\text{D}},i})$ are projectors onto arbitrary single-ququart basis. The measurement result on $\ket{g(\vec{\theta}_{\text{G}})}$ can be estimated from the probability that the signal and idler photons are coincidentally detected at both second ports \cite{Wang2023}.

To implement the QGAN, we use the two-ququart maximally-entangled state preparation parts to act as the \emph{Generator} circuit that can produce the candidate quantum state $\hat{\rho}(\vec{\theta}_{\text{G}})=\ket{g(\vec{\theta}_{\text{G}})}\bra{g(\vec{\theta}_{\text{G}})}$, and the two-ququart measurement part as the \emph{Discriminator} circuit where the measurement operator can be described by $\mathcal{\hat{M}}(\vec{\theta}_{\text{D}})$. 
In our training process, the two players utilize the gradient descent method to update their parameters. The partial derivatives of the measurement difference $\text{d}(\vec{\theta}_{\text{G}},\vec{\theta}_{\text{D}})$ with respect to  $\vec{\theta}_{\text{G}}$ and $\vec{\theta}_{\text{D}}$ can be estimated using the parameter-shift rule \cite{Wang2023}. See Supplement 1 for more experimental details.

\begin{figure}[!t]
  \centering\includegraphics{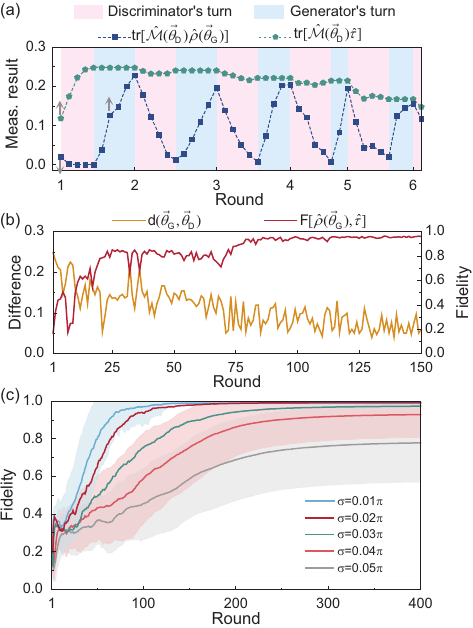}
  \caption{Quantum generative adversarial learning using photonic circuits in the presence of noise. 
  (a) A comparison of the measurement results between the generated state $\text{tr}[\mathcal{\hat{M}}(\vec{\theta}_{\text{D}})\hat{\rho}(\vec{\theta}_{\text{G}})]$ and the true state $\text{tr}[\mathcal{\hat{M}}(\vec{\theta}_{\text{D}})\hat{\tau}]$ in the first five rounds of the the QGAN experiment.
  (b) Evolution of the generated state fidelity $\text{F}[\hat{\rho}(\vec{\theta}_{\text{G}}),\hat{\tau}]$ and the measurement difference $\text{d}(\vec{\theta}_{\text{G}},\vec{\theta}_{\text{D}})$.
  (c) The convergence process of the generated state fidelity with QGANs subjected to different levels of phase noise. For every noise level, the average and standard deviation (shaded area, except $\sigma=0.2,0.3$ for simplicity) of the fidelities in 100 simulations are plotted. 
  } 
  \label{fig: Full paras}
\end{figure}
\textit{Quantum generative adversarial learning in circuits subjected to noise}.—
In this section, QGANs are initialized as follows: 
The true data $\hat{\tau}$ is a maximally entangled state of two ququarts that is randomly generated.
The generator utilizes two intact universal linear optical circuits and initially sets the single-ququart operations $\hat{U}_s(\vec{\theta}_s)$ and $\hat{U}_i(\vec{\theta}_i)$ to implement identity operations. For the discriminator, the two single-ququart projectors $\hat{P}_s(\vec{\theta}_{{\text{D}},s})$ and $\hat{P}_i(\vec{\theta}_{{\text{D}},i})$ that form the measurement $\mathcal{\hat{M}}(\vec{\theta}_{\text{D}})$ are also randomly initialized. 

We first perform experimental demonstrations using our photonic chip, which is a practical noisy device illustrated schematically in Fig. \ref{fig: setup}(e). 
Fig. \ref{fig: Full paras}(a) provides details regarding the evolution of measurement results $\text{tr}[\mathcal{\hat{M}}(\vec{\theta}_{\text{D}})\hat{\rho}(\vec{\theta}_{\text{G}})]$ and $\text{tr}[\mathcal{\hat{M}}(\vec{\theta}_{\text{D}})\hat{\tau}]$ in the first five rounds of the adversarial learning.
It clearly shows that in discriminator's turn, the updated measurement $\mathcal{\hat{M}}(\vec{\theta}_{\text{D}})$ leads to an increasing gap in the measurement results between the generated state $\hat{\rho}(\vec{\theta}_{\text{G}})$ and the true state $\hat{\tau}$, while in generator's turn, the measurement results of $\hat{\rho}(\vec{\theta}_{\text{G}})$ approach those of $\hat{\tau}$, and the difference narrows.
Fig. \ref{fig: Full paras}(b) shows the whole convergence process of the measurement difference $\text{d}(\vec{\theta}_{\text{G}},\vec{\theta}_{\text{D}})$ and the quantum state fidelities $\text{F}[\hat{\rho}(\vec{\theta}_{\text{G}}),\hat{\tau}] = \text{tr}[\sqrt{\hat{\rho}^{1/2}(\vec{\theta}_{\text{G}})\hat{\tau}\hat{\rho}^{1/2}(\vec{\theta}_{\text{G}})}]$ \cite{Nielsen2010}. 
After 150 rounds of competitions, although the measurement difference $\text{d}(\vec{\theta}_{\text{G}},\vec{\theta}_{\text{D}})$ converges to about 0.05 rather than 0 due to noise, the fidelity of the generated state increases from $16.96\%$ to $96.02\%$. 

As a supplement to our experiments, numerical simulations have been performed to demonstrate the robustness of QGANs against noise. 
Two types of noise are considered: phase noise, whose intensity is described by $\sigma$, and shot noise, whose intensity is described by the total two-photon coincidence count $\Vert \vec{c} \Vert$.
In our simulations, we set $\Vert \vec{c} \Vert =1500$, which is the same in our experiments. 
With $\sigma$ ranging from $0.01\pi$ to $0.05\pi$, we performed 100 simulations for each phase noise level. 
As shown in Fig. \ref{fig: Full paras}(c), when the phase noise level $\sigma$ is less than $0.04\pi$, the final fidelity is expected to exceed $90\%$. However, when $\sigma$ reaches $0.05\pi$, the fidelity drops below $80\%$. Note that $0.04\pi$ is a value much higher than the actual phase noise level in our chip.
These results demonstrate the ability of QGANs to generate high-quality quantum states on near-term noisy devices.

\begin{figure}[!t]
  \centering\includegraphics{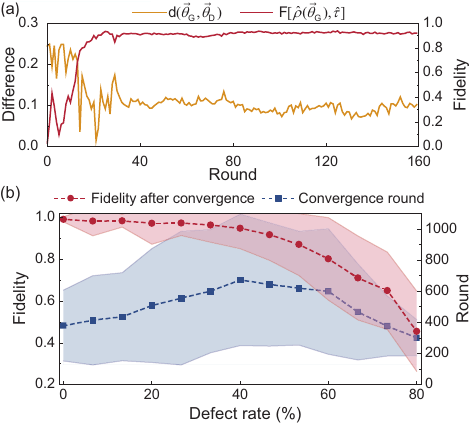}
  \caption{Quantum generative adversarial learning in circuits with defects. 
  (a) The evolution of generated state fidelity $\text{F}[\hat{\rho}(\vec{\theta}_{\text{G}}),\hat{\tau}]$ and measurement difference $\text{d}(\vec{\theta}_{\text{G}},\vec{\theta}_{\text{D}})$ in QGAN training on defective circuits. In this experimental demonstration, 8 out of 15 phase shifters in each of the two universal linear circuits of the generator were set to fixed values and marked with triangle symbols in Fig. \ref{fig: setup}(f). After around 160 rounds, the fidelity has risen to $92.11\%$.
  (b) The fidelity and number of learning rounds after successful convergence with varying defect rates in the generator's circuit. Here, defect rate is the total percentage of the defective phase shifters in the generator's circuits. In each trial, the defective phase shifters were randomly selected. Even up to half of generator's phase shifters are damaged, the generated state fidelity can be expected to higher than $90\%$.}
  \label{fig: Defective chip}
\end{figure}

\textit{Quantum generative adversarial learning in defective circuits}.—\
Generally, defective quantum devices are kept in use by excluding the defective parts, resulting in a reduction of their usable size \cite{Arute2019,Gong2021,Wang2018}. 
In this section, we confirmed through experiments and numerical simulations that QGANs can be applied to prepare high-fidelity quantum states on defective chips instead of reducing their scale, even if more than half of the necessary phase shifters on the quantum photonic chip are out of control. 
Reviewing Fig. \ref{fig: setup}(d), the 4-mode universal linear optical circuit \cite{Clements2016} in our chip comprises 15 programmable phase shifters, which has the minimum number of degrees of freedom required to achieve arbitrary 4-dimensional unitary transformations. 
We experimentally investigated quantum generative adversarial learning under the condition that 16 of the 30 generator's phase shifters were defective. 
In Fig. \ref{fig: setup}(f), 8 phase shifters assumed to be defective in each of the two universal linear circuits are set as some random and fixed values and marked with triangle symbols.
As shown in Fig. \ref{fig: Defective chip}(a), the learning process converges after approximately 160 rounds, with the fidelity rising to $92.11\%$. Compared with the learning results in intact circuits (Fig. \ref{fig: Full paras}(b)), defects reduce the final fidelity from $96.02\%$ to $92.11\%$, while the final measurement difference increases from $\sim0.05$ to $\sim0.10$. 
These results show that although severe defects decrease the quality of generated quantum data, QGAN can find a solution to generate quantum data with fidelity higher than $90\%$ even when up to half of the generator's phase shifters are damaged.

In addition to our experiment, we performed 100 simulations for each defective rate increasing from 0 to $80\%$.
In each simulation, the broken phase shifters were randomly selected and set to fixed values. Phase noise ($\sigma=0.02\pi$) and shot noise ($\Vert \vec{c} \Vert =1500$) were taken into consideration. 
Fig. \ref{fig: Defective chip}(b) shows the downtrend of the final fidelities with the increasing defects. At first, the fidelity can reach almost up to $100\%$ in intact circuits. Then the convergence cost gradually increases as more defects are added, due to the increasing difficulty of finding solutions in worse circuit conditions. 
Even if approximately half of the phase shifters are still functioning correctly, the fidelity of the quantum state will exceed $90\%$, as we achieved in the experiment.
Finally, the continuous deterioration of circuits leads to a lower achievable fidelity and an earlier convergence of training rounds. 
These results show that as long as half of the generator's phase shifters are defective, QGANs can still work normally on our photonic chip and generate high-quality quantum data.

\textit{Conclusion}.—
We have used a programmable silicon quantum photonic chip to implement QGANs experimentally in photonics for the first time. 
Our results demonstrate the feasibility of QGANs on near-term quantum photonic devices, as well as their robustness against noise and defects.
The future large-scale quantum photonic circuits with more photons will face complex noise and defects, and their effects on QGANs warrant further study.
The methods used to generate quantum data with QGANs are ready to be applied to larger-sized systems with more photons and different circuits in photonics and on various physical platforms.
We would also like to point out that the circuits for the quantum generator and discriminator do not demand for the implementation of universal quantum logic or rigorous fabrication quality. This indicates that QGANs can be achieved on near-term and imperfect quantum devices, offering possibilities for achieving quantum advantages and practical applications in the NISQ era.

\begin{backmatter}
\bmsection{Funding} National Natural Science Foundation of China (62061136011, 62105366, and 62075243); Aid Program for Science and Technology Innovative Research Team in Higher Educational Institutions of Hunan Province.


\bmsection{Disclosures} The authors declare no conflicts of interest.

\bmsection{Data availability} Data underlying the results presented in this paper are not publicly available at this time but may be obtained from the authors upon reasonable request.

\bmsection{Supplemental document}
See Supplementary document for supporting content.

\end{backmatter}

\bibliography{ms}

\begin{thebibliography}{10}
\newcommand{\enquote}[1]{``#1''}

\bibitem{Nielsen2010}
M.~A. Nielsen and I.~L. Chuang, \emph{Quantum Computation and Quantum
  Information} ({Cambridge University Press}, {Cambridge ; New York}, 2010),
  10th ed.

\bibitem{Cerezo2022}
M.~Cerezo, G.~Verdon, H.-Y. Huang, L.~Cincio, and P.~J. Coles,
  {\protect\JournalTitle{Nature Computational Science}} \textbf{2}, 567 (2022).

\bibitem{Lloyd2018a}
S.~Lloyd and C.~Weedbrook, {\protect\JournalTitle{Physical Review Letters}}
  \textbf{121}, 040502 (2018).

\bibitem{Dallaire-Demers2018}
P.-L. {Dallaire-Demers} and N.~Killoran, {\protect\JournalTitle{Physical Review
  A}} \textbf{98}, 012324 (2018).

\bibitem{Goodfellow2014}
I.~J. Goodfellow, J.~Pouget-Abadie, M.~Mirza, B.~Xu, D.~Warde-Farley, S.~Ozair,
  A.~Courville, and Y.~Bengio, \enquote{Generative adversarial nets,} in
  \emph{Proceedings of the 27th International Conference on Neural Information
  Processing Systems,}  (MIT Press, Cambridge, MA, USA, 2014).

\bibitem{Zoufal2019}
C.~Zoufal, A.~Lucchi, and S.~Woerner, {\protect\JournalTitle{npj Quantum
  Information}} \textbf{5}, 1 (2019).

\bibitem{Situ2020}
H.~Situ, Z.~He, Y.~Wang, l.~Li, and S.~Zheng,
  {\protect\JournalTitle{Information Sciences}} \textbf{538}, 193 (2020).

\bibitem{Romero2021}
J.~Romero and A.~{Aspuru-Guzik}, {\protect\JournalTitle{Advanced Quantum
  Technologies}} \textbf{4}, 2000003 (2021).

\bibitem{Huang2021a}
H.-L. Huang, Y.~Du, M.~Gong, Y.~Zhao, Y.~Wu, C.~Wang, S.~Li, F.~Liang, J.~Lin,
  Y.~Xu, R.~Yang, T.~Liu, M.-H. Hsieh, H.~Deng, H.~Rong, C.-Z. Peng, C.-Y. Lu,
  Y.-A. Chen, D.~Tao, X.~Zhu, and J.-W. Pan, {\protect\JournalTitle{Physical
  Review Applied}} \textbf{16}, 024051 (2021).

\bibitem{Li2021a}
J.~Li, R.~O. Topaloglu, and S.~Ghosh, {\protect\JournalTitle{IEEE Transactions
  on Quantum Engineering}} \textbf{2}, 1 (2021).

\bibitem{Hu2019}
L.~Hu, S.-H. Wu, W.~Cai, Y.~Ma, X.~Mu, Y.~Xu, H.~Wang, Y.~Song, D.-L. Deng,
  C.-L. Zou, and L.~Sun, {\protect\JournalTitle{Science Advances}} \textbf{5},
  eaav2761 (2019).

\bibitem{Huang2021}
K.~Huang, Z.-A. Wang, C.~Song, K.~Xu, H.~Li, Z.~Wang, Q.~Guo, Z.~Song, Z.-B.
  Liu, D.~Zheng, D.-L. Deng, H.~Wang, J.-G. Tian, and H.~Fan,
  {\protect\JournalTitle{npj Quantum Information}} \textbf{7}, 165 (2021).

\bibitem{Ahmed2021}
S.~Ahmed, C.~S{\'a}nchez~Mu{\~n}oz, F.~Nori, and A.~F. Kockum,
  {\protect\JournalTitle{Physical Review Letters}} \textbf{127}, 140502 (2021).

\bibitem{Anand2021a}
A.~Anand, J.~Romero, M.~Degroote, and A.~Aspuru-Guzik,
  {\protect\JournalTitle{Advanced Quantum Technologies}} \textbf{4}, 2000069
  (2021).

\bibitem{Niu2022b}
M.~Y. Niu, A.~Zlokapa, M.~Broughton, S.~Boixo, M.~Mohseni, V.~Smelyanskyi, and
  H.~Neven, {\protect\JournalTitle{Physical Review Letters}} \textbf{128},
  220505 (2022).

\bibitem{Zhu2022}
E.~Y. Zhu, S.~Johri, D.~Bacon, M.~Esencan, J.~Kim, M.~Muir, N.~Murgai,
  J.~Nguyen, N.~Pisenti, A.~Schouela, K.~Sosnova, and K.~Wright,
  {\protect\JournalTitle{Physical Review Research}} \textbf{4}, 043092 (2022).

\bibitem{Preskill2018}
J.~Preskill, {\protect\JournalTitle{Quantum}} \textbf{2}, 79 (2018).

\bibitem{Arute2019}
{Google AI Quantum and Collaborators}, {\protect\JournalTitle{Nature}}
  \textbf{574}, 505 (2019).

\bibitem{Gong2021}
M.~Gong, S.~Wang, C.~Zha, M.-C. Chen, H.-L. Huang, Y.~Wu, Q.~Zhu, Y.~Zhao,
  S.~Li, S.~Guo, H.~Qian, Y.~Ye, F.~Chen, C.~Ying, J.~Yu, D.~Fan, D.~Wu, H.~Su,
  H.~Deng, H.~Rong, K.~Zhang, S.~Cao, J.~Lin, Y.~Xu, L.~Sun, C.~Guo, N.~Li,
  F.~Liang, V.~M. Bastidas, K.~Nemoto, W.~J. Munro, Y.-H. Huo, C.-Y. Lu, C.-Z.
  Peng, X.~Zhu, and J.-W. Pan, {\protect\JournalTitle{Science}} \textbf{372},
  948 (2021).

\bibitem{Wang2018}
J.~Wang, S.~Paesani, Y.~Ding, R.~Santagati, P.~Skrzypczyk, A.~Salavrakos,
  J.~Tura, R.~Augusiak, L.~Man{\v c}inska, D.~Bacco, D.~Bonneau, J.~W.
  Silverstone, Q.~Gong, A.~Ac{\'i}n, K.~Rottwitt, L.~K. Oxenl{\o}we, J.~L.
  O'Brien, A.~Laing, and M.~G. Thompson, {\protect\JournalTitle{Science}}
  \textbf{360}, 285 (2018).

\bibitem{Qiang2021}
X.~Qiang, Y.~Wang, S.~Xue, R.~Ge, L.~Chen, Y.~Liu, A.~Huang, X.~Fu, P.~Xu,
  T.~Yi, F.~Xu, M.~Deng, J.~B. Wang, J.~D.~A. Meinecke, J.~C.~F. Matthews,
  X.~Cai, X.~Yang, and J.~Wu, {\protect\JournalTitle{Science Advances}}
  \textbf{7}, eabb8375 (2021).

\bibitem{Nashequilibrium}
J.~F. Nash, {\protect\JournalTitle{Proceedings of the National Academy of
  Sciences}} \textbf{36}, 48 (1950).

\bibitem{Borras2023}
K.~Borras, S.~Y. Chang, L.~Funcke, M.~Grossi, T.~Hartung, K.~Jansen,
  D.~Kruecker, S.~K{\"u}hn, F.~Rehm, C.~T{\"u}ys{\"u}z, and S.~Vallecorsa,
  {\protect\JournalTitle{Journal of Physics: Conference Series}} \textbf{2438},
  012093 (2023).

\bibitem{Paesani2017a}
S.~Paesani, A.~A. Gentile, R.~Santagati, J.~Wang, N.~Wiebe, D.~P. Tew, J.~L.
  O'Brien, and M.~G. Thompson, {\protect\JournalTitle{Physical Review Letters}}
  \textbf{118}, 100503 (2017).

\bibitem{Adcock2021}
J.~C. Adcock, J.~Bao, Y.~Chi, X.~Chen, D.~Bacco, Q.~Gong, L.~K. Oxenlowe,
  J.~Wang, and Y.~Ding, {\protect\JournalTitle{IEEE Journal of Selected Topics
  in Quantum Electronics}} \textbf{27}, 1 (2021).

\bibitem{Xue2022}
S.~Xue, Y.~Wang, J.~Zhan, Y.~Wang, R.~Zeng, J.~Ding, W.~Shi, Y.~Liu, Y.~Liu,
  A.~Huang, G.~Huang, C.~Yu, D.~Wang, X.~Fu, X.~Qiang, P.~Xu, M.~Deng, X.~Yang,
  and J.~Wu, {\protect\JournalTitle{Physical Review Letters}} \textbf{129},
  133601 (2022).

\bibitem{Wang2023}
Y.~Wang, S.~Xue, Y.~Wang, J.~Ding, W.~Shi, D.~Wang, Y.~Liu, Y.~Liu, X.~Fu,
  G.~Huang, A.~Huang, M.~Deng, and J.~Wu, {\protect\JournalTitle{Optics
  Letters}} \textbf{48}, 3745 (2023).

\bibitem{Silverstone2014}
J.~W. Silverstone, D.~Bonneau, K.~Ohira, N.~Suzuki, H.~Yoshida, N.~Iizuka,
  M.~Ezaki, C.~M. Natarajan, M.~G. Tanner, R.~H. Hadfield, V.~Zwiller, G.~D.
  Marshall, J.~G. Rarity, J.~L. O'Brien, and M.~G. Thompson,
  {\protect\JournalTitle{Nature Photonics}} \textbf{8}, 104 (2014).

\bibitem{Clements2016}
W.~R. Clements, P.~C. Humphreys, B.~J. Metcalf, W.~S. Kolthammer, and I.~A.
  Walsmley, {\protect\JournalTitle{Optica}} \textbf{3}, 1460 (2016).

\end{thebibliography}


\begin{thebibliography}{10}
\newcommand{\enquote}[1]{``#1''}

\bibitem{Xue2022}
S.~Xue, Y.~Wang, J.~Zhan, Y.~Wang, R.~Zeng, J.~Ding, W.~Shi, Y.~Liu, Y.~Liu,
  A.~Huang, G.~Huang, C.~Yu, D.~Wang, X.~Fu, X.~Qiang, P.~Xu, M.~Deng, X.~Yang,
  and J.~Wu, \enquote{Variational {{Entanglement-Assisted Quantum Process
  Tomography}} with {{Arbitrary Ancillary Qubits}},}
  {\protect\JournalTitle{Physical Review Letters}} \textbf{129}, 133601 (2022).

\bibitem{Wang2023}
Y.~Wang, S.~Xue, Y.~Wang, J.~Ding, W.~Shi, D.~Wang, Y.~Liu, Y.~Liu, X.~Fu,
  G.~Huang, A.~Huang, M.~Deng, and J.~Wu, \enquote{Experimental quantum natural
  gradient optimization in photonics,} {\protect\JournalTitle{Optics Letters}}
  \textbf{48}, 3745 (2023).

\bibitem{Silverstone2014}
J.~W. Silverstone, D.~Bonneau, K.~Ohira, N.~Suzuki, H.~Yoshida, N.~Iizuka,
  M.~Ezaki, C.~M. Natarajan, M.~G. Tanner, R.~H. Hadfield, V.~Zwiller, G.~D.
  Marshall, J.~G. Rarity, J.~L. O'Brien, and M.~G. Thompson, \enquote{On-chip
  quantum interference between silicon photon-pair sources,}
  {\protect\JournalTitle{Nature Photonics}} \textbf{8}, 104--108 (2014).

\bibitem{Liu2020}
Y.~Liu, C.~Wu, X.~Gu, Y.~Kong, X.~Yu, R.~Ge, X.~Cai, X.~Qiang, J.~Wu, X.~Yang,
  and P.~Xu, \enquote{High-spectral-purity photon generation from a
  dual-interferometer-coupled silicon microring,} {\protect\JournalTitle{Optics
  Letters}} \textbf{45}, 73 (2020).

\bibitem{Clements2016}
W.~R. Clements, P.~C. Humphreys, B.~J. Metcalf, W.~S. Kolthammer, and I.~A.
  Walsmley, \enquote{Optimal design for universal multiport interferometers,}
  {\protect\JournalTitle{Optica}} \textbf{3}, 1460 (2016).

\bibitem{Arrazola2021}
J.~M. Arrazola, V.~Bergholm, K.~Br{\'a}dler, T.~R. Bromley, M.~J. Collins,
  I.~Dhand, A.~Fumagalli, T.~Gerrits, A.~Goussev, L.~G. Helt, J.~Hundal,
  T.~Isacsson, R.~B. Israel, J.~Izaac, S.~Jahangiri, R.~Janik, N.~Killoran,
  S.~P. Kumar, J.~Lavoie, A.~E. Lita, D.~H. Mahler, M.~Menotti, B.~Morrison,
  S.~W. Nam, L.~Neuhaus, H.~Y. Qi, N.~Quesada, A.~Repingon, K.~K. Sabapathy,
  M.~Schuld, D.~Su, J.~Swinarton, A.~Sz{\'a}va, K.~Tan, P.~Tan, V.~D. Vaidya,
  Z.~Vernon, Z.~Zabaneh, and Y.~Zhang, \enquote{Quantum circuits with many
  photons on a programmable nanophotonic chip,} {\protect\JournalTitle{Nature}}
  \textbf{591}, 54--60 (2021).

\bibitem{Chi2022}
Y.~Chi, J.~Huang, Z.~Zhang, J.~Mao, Z.~Zhou, X.~Chen, C.~Zhai, J.~Bao, T.~Dai,
  H.~Yuan, M.~Zhang, D.~Dai, B.~Tang, Y.~Yang, Z.~Li, Y.~Ding, L.~K.
  Oxenl{\o}we, M.~G. Thompson, J.~L. O'Brien, Y.~Li, Q.~Gong, and J.~Wang,
  \enquote{A programmable qudit-based quantum processor,}
  {\protect\JournalTitle{Nature Communications}} \textbf{13}, 1166 (2022).

\bibitem{Taballione2023}
C.~Taballione, M.~C. Anguita, M.~De~Goede, P.~Venderbosch, B.~Kassenberg,
  H.~Snijders, N.~Kannan, W.~L. Vleeshouwers, D.~Smith, J.~P. Epping, R.~Van
  Der~Meer, P.~W.~H. Pinkse, H.~Van Den~Vlekkert, and J.~J. Renema,
  \enquote{20-{{Mode Universal Quantum Photonic Processor}},}
  {\protect\JournalTitle{Quantum}} \textbf{7}, 1071 (2023).

\bibitem{Nielsen2010}
M.~A. Nielsen and I.~L. Chuang, \emph{Quantum Computation and Quantum
  Information} ({Cambridge University Press}, {Cambridge ; New York}, 2010),
  10th ed.

\bibitem{Wang2018}
J.~Wang, S.~Paesani, Y.~Ding, R.~Santagati, P.~Skrzypczyk, A.~Salavrakos,
  J.~Tura, R.~Augusiak, L.~Man{\v c}inska, D.~Bacco, D.~Bonneau, J.~W.
  Silverstone, Q.~Gong, A.~Ac{\'i}n, K.~Rottwitt, L.~K. Oxenl{\o}we, J.~L.
  O'Brien, A.~Laing, and M.~G. Thompson, \enquote{Multidimensional quantum
  entanglement with large-scale integrated optics,}
  {\protect\JournalTitle{Science}} \textbf{360}, 285--291 (2018).

\bibitem{Vigliar2021}
C.~Vigliar, S.~Paesani, Y.~Ding, J.~C. Adcock, J.~Wang, S.~{Morley-Short},
  D.~Bacco, L.~K. Oxenl{\o}we, M.~G. Thompson, J.~G. Rarity, and A.~Laing,
  \enquote{Error-protected qubits in a silicon photonic chip,}
  {\protect\JournalTitle{Nature Physics}} \textbf{17}, 1137--1143 (2021).

\bibitem{Bao2023}
J.~Bao, Z.~Fu, T.~Pramanik, J.~Mao, Y.~Chi, Y.~Cao, C.~Zhai, Y.~Mao, T.~Dai,
  X.~Chen, X.~Jia, L.~Zhao, Y.~Zheng, B.~Tang, Z.~Li, J.~Luo, W.~Wang, Y.~Yang,
  Y.~Peng, D.~Liu, D.~Dai, Q.~He, A.~L. Muthali, L.~K. Oxenl{\o}we, C.~Vigliar,
  S.~Paesani, H.~Hou, R.~Santagati, J.~W. Silverstone, A.~Laing, M.~G.
  Thompson, J.~L. O'Brien, Y.~Ding, Q.~Gong, and J.~Wang,
  \enquote{Very-large-scale integrated quantum graph photonics,}
  {\protect\JournalTitle{Nature Photonics}}  (2023).

\bibitem{Qiang2018}
X.~Qiang, X.~Zhou, J.~Wang, C.~M. Wilkes, T.~Loke, S.~O'Gara, L.~Kling, G.~D.
  Marshall, R.~Santagati, T.~C. Ralph, J.~B. Wang, J.~L. O'Brien, M.~G.
  Thompson, and J.~C.~F. Matthews, \enquote{Large-scale silicon quantum
  photonics implementing arbitrary two-qubit processing,}
  {\protect\JournalTitle{Nature Photonics}} \textbf{12}, 534--539 (2018).

\bibitem{Qiang2021}
X.~Qiang, Y.~Wang, S.~Xue, R.~Ge, L.~Chen, Y.~Liu, A.~Huang, X.~Fu, P.~Xu,
  T.~Yi, F.~Xu, M.~Deng, J.~B. Wang, J.~D.~A. Meinecke, J.~C.~F. Matthews,
  X.~Cai, X.~Yang, and J.~Wu, \enquote{Implementing graph-theoretic quantum
  algorithms on a silicon photonic quantum walk processor,}
  {\protect\JournalTitle{Science Advances}} \textbf{7}, eabb8375 (2021).

\bibitem{Wang2020a}
J.~Wang, F.~Sciarrino, A.~Laing, and M.~G. Thompson, \enquote{Integrated
  photonic quantum technologies,} {\protect\JournalTitle{Nature Photonics}}
  \textbf{14}, 273--284 (2020).

\bibitem{Gilardi2014}
G.~Gilardi, {Weiming Yao}, H.~Rabbani~Haghighi, X.~J.~M. Leijtens, M.~K. Smit,
  and M.~J. Wale, \enquote{Deep {{Trenches}} for {{Thermal Crosstalk
  Reduction}} in {{InP-Based Photonic Integrated Circuits}},}
  {\protect\JournalTitle{Journal of Lightwave Technology}} \textbf{32},
  4864--4870 (2014).

\bibitem{Milanizadeh2019}
M.~Milanizadeh, D.~Aguiar, A.~Melloni, and F.~Morichetti, \enquote{Canceling
  {{Thermal Cross-Talk Effects}} in {{Photonic Integrated Circuits}},}
  {\protect\JournalTitle{Journal of Lightwave Technology}} \textbf{37},
  1325--1332 (2019).

\bibitem{De2020}
S.~De, R.~Das, R.~K. Varshney, and T.~Schneider, \enquote{Design and
  {{Simulation}} of {{Thermo-Optic Phase Shifters With Low Thermal Crosstalk}}
  for {{Dense Photonic Integration}},} {\protect\JournalTitle{IEEE Access}}
  \textbf{8}, 141632--141640 (2020).

\bibitem{Paesani2017a}
S.~Paesani, A.~A. Gentile, R.~Santagati, J.~Wang, N.~Wiebe, D.~P. Tew, J.~L.
  O'Brien, and M.~G. Thompson, \enquote{Experimental {{Bayesian Quantum Phase
  Estimation}} on a {{Silicon Photonic Chip}},} {\protect\JournalTitle{Physical
  Review Letters}} \textbf{118}, 100503 (2017).

\bibitem{Adcock2021}
J.~C. Adcock, J.~Bao, Y.~Chi, X.~Chen, D.~Bacco, Q.~Gong, L.~K. Oxenlowe,
  J.~Wang, and Y.~Ding, \enquote{Advances in {{Silicon Quantum Photonics}},}
  {\protect\JournalTitle{IEEE Journal of Selected Topics in Quantum
  Electronics}} \textbf{27}, 1--24 (2021).

\bibitem{Campbell1909a}
N.~Campbell, \enquote{The study of discontinuous phenomena,}
  {\protect\JournalTitle{Proc. Cambr. Phil. Soc.}} \textbf{15}, 117--136
  (1909).

\bibitem{Campbell1909b}
N.~Campbell, \enquote{Discontinuities in light emission,}
  {\protect\JournalTitle{Proc. Cambr. Phil. Soc.}} \textbf{15}, 310--328
  (1909).

\bibitem{Eliazar2005}
I.~Eliazar and J.~Klafter, \enquote{On the nonlinear modeling of shot noise,}
  {\protect\JournalTitle{Proceedings of the National Academy of Sciences}}
  \textbf{102}, 13779--13782 (2005).

\end{thebibliography}




\end{document}


\maketitle

\section{Photonic chip structure for the QGAN model}

\begin{figure*}[!h]
  \centering\includegraphics[width=18cm]{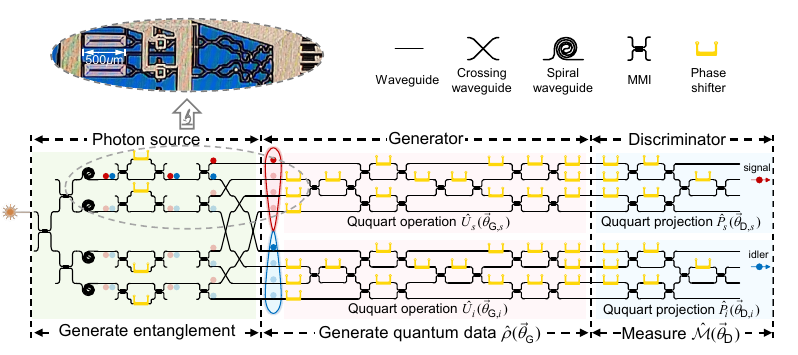}
  \caption{Schematic of the reconfigurable silicon photonic chip used in the quantum generative adversarial learning experiments. The chip includes three functional parts: (i) on-chip photon pair sources that generate two-ququart entangled states, (ii) universal linear optical circuits to implement arbitrary unitary operations on each ququart, and (iii) measurement circuits to perform arbitrary projections on each ququart. Note that three external phase shifters located at the forepart of each projection circuit are not included in the manufactured chip. We use phase shifters located at the rear of the universal linear optical circuits to obtain the necessary additional phase shifts for measurements. Finally, the chip structure monolithically embeds four spontaneous four-wave mixing photon-pair sources, 40 reconfigurable thermo-optic phase shifters, and 44 multimode interferometers. The inset shows a micrograph that captures all the critical components of the chip.}
  \label{fig: chip}
\end{figure*}

We use a reconfigurable silicon photonic chip that has been reported in Refs. \cite{Xue2022,Wang2023} to perform experimental quantum generative adversarial learning demonstrations. As shown in Fig. \ref{fig: chip}, the chip includes three functional parts: (i) on-chip photon pair sources that generate two-ququart entangled states, (ii) universal linear optical circuits to implement arbitrary unitary operations on each ququarts, and (iii) measurement circuits to perform arbitrary projections on each ququart.

\subsection{Source: generating the two-photon maximally entangled state}

In the part of entangled photon generation, the chip contains a tree-organized beam splitter network, four
spontaneous four-wave mixing (SFWM) photon-pair sources \cite{Silverstone2014}, and four asymmetric Mach-Zehnder interferometers (AMZIs) \cite{Liu2020}.
As shown in the micrograph of Fig. \ref{fig: chip}, the four photon-pair sources were designed with the same spiral waveguide length of 1.5 cm. The SFWM nonlinear effect in silicon is exploited to produce non-degenerate signal-idler photon pairs \cite{Silverstone2014}.
The 50:50 beam splitters distribute the input pump laser equally, and the four SFWM sources are coherently pumped.
By configuring the following four AMZIs, the photon pairs from all four sources can be retained while the residual pump laser is filtered out of the chip. 
Post-selecting the cases in which signal photons exit from the top circuit and idler photons exit from the bottom circuit yields a two-photon maximally entangled state
\begin{equation}
  \ket{\psi_0}=\frac{1}{2}\sum_{k=1}^4 {\ket{k}_s\otimes\ket{k}_i}.
\end{equation}
Here, $\ket{k}_s$ and $\ket{k}_i$ represent the path modes of the signal and idler photons.

\subsection{Generator's circuit: implementing arbitrary unitary operations on each ququart}

\begin{figure*}[!h]
  \centering\includegraphics{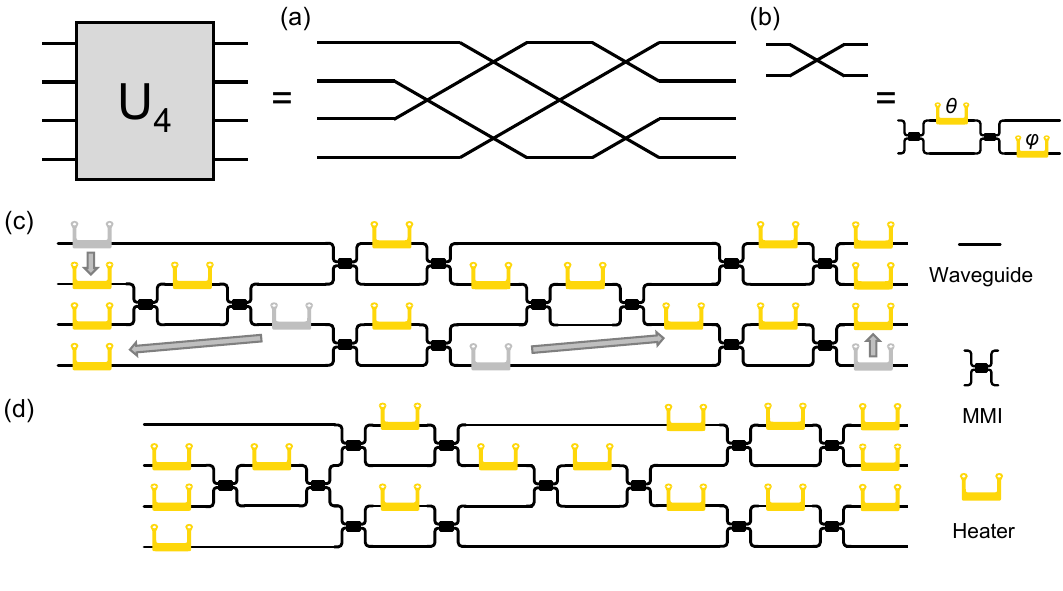}
  \caption{Schematic of the 4-mode universal linear optical circuits.
  (a) A universal 4-mode unitary operation can be implemented using a mesh of six beam splitters.
  (b) A line corresponds to an optical mode, and crossings between two modes correspond to two 2$\times$2 MMIs, followed by a phase shifter at one output port.
  (c) The complete realization of (a).
  (d) The final realization of the square mesh on our chip was achieved by reducing some redundant phase shifters, which were colored in gray in (c).
  }
  \label{fig: U}
\end{figure*}

One of the basic building blocks for integrated photonic circuits is the beam splitter, equivalent to classical beam splitter that splits power across two paths.  We use the multi-mode interferometer (MMI) in our chip to implement the 2$\times$2 beam splitter transformation
\begin{equation}
  \mathrm{BS} = \frac{1}{\sqrt{2}}
  \begin{bmatrix}
      1 & i \\
      \mathrm{{i}} & 1
  \end{bmatrix}.
\end{equation}
The structure of a 2$\times$2 MMI consists of a multimode waveguide and two single-mode waveguides connected to the multimode waveguide at both sides as two inputs and two outputs, respectively. Phase shifter is another essential building block that is the only variable component of an integrated photonic circuit.  In our chip, phase shifter is a phase gate used to induce a programmable relative phase difference $\theta$ between photon paths by thermo-optic effects. For simplicity, we describe the operation of phase shifter $\mathrm{PS}(\theta)$ in a 2-dimensional Hilbert space as
\begin{equation}\label{Eq: PS}
  \mathrm{PS}(\theta) = 
      \begin{bmatrix}
          \mathrm{e}^{i\theta} & 0 \\
          0 & 1
      \end{bmatrix}.
\end{equation}
With two balanced beam splitters and an internal phase shifter on one of the two paths between the beam splitters, a Mach-Zehnder interferometer (MZI) can perform the transformation:
\begin{equation}
    \textrm{MZI}(\theta)=\textrm{BS}\cdot\textrm{PS}(\theta)\cdot\textrm{BS}=\begin{bmatrix}
        i\mathrm{e}^{i\frac{\theta}{2}}\sin\frac{\theta}{2} & i\mathrm{e}^{i\frac{\theta}{2}}\cos\frac{\theta}{2}\\
        i\mathrm{e}^{i\frac{\theta}{2}}\cos\frac{\theta}{2} & -i\mathrm{e}^{i\frac{\theta}{2}}\sin\frac{\theta}{2}
    \end{bmatrix}.
\end{equation}

Following Ref. \cite{Clements2016}, MZIs and extra phase shifters arranged in a square mesh can be programmed to implement any linear transformation matrix \cite{Arrazola2021,Chi2022,Taballione2023}. The square mesh structure for the implementation of four-dimensional unitary operations is shown in Fig. \ref{fig: U}(a). The basic module comprises an MZI and a following phase shifter at one output port. The scheme by Ref. \cite{Clements2016} is based on analytical methods of decomposing the U matrix into a product of basic module operations
\begin{equation}
  \hat{U}=\hat{D}\prod_{k=1}\hat{T}_k(\theta_k,\phi_k),
\end{equation}
where $\hat{T}(\theta,\phi)$ is the operation of a basic module in Fig. \ref{fig: U}(b) with values $\theta$ and $\phi$ defined by the internal and external phase shifters, and $\hat{D}$ is a diagonal matrix about the relative phase of the input and output modes.

The complete realization of the square mesh preceded and tailed by extra phase shifters (used for adjusting the relative phase of the input and output modes) is presented in Fig. \ref{fig: U}(c). 
Fig. \ref{fig: U}(d) presents the final realization of the square mesh on our chip. We reduce some redundant phase shifters that are colored in gray in Fig. \ref{fig: U}(c) by merging two neighboring phase shifters.
With the structure in Fig. \ref{fig: U}(d), we can decompose an arbitrary 4-mode $U$ matrix by programming the 15 phase shifters.

Using the two 4-mode universal linear optical circuits of the chip, the generator can perform arbitrary single-ququart unitary transformations, $\hat{U}_s(\vec{\theta}_{{\text{G}},s})$ and $\hat{U}_i(\vec{\theta}_{{\text{G}},i})$, on the signal and idler of the two photons. Then the evolved two-photon state becomes
\begin{equation} \label{Eq: Schmidt decomposition}
  \ket{g(\vec{\theta}_{\text{G}})}=\frac{1}{2} \hat{U}_s(\vec{\theta}_{{\text{G}},s}) \otimes \hat{U}_i(\vec{\theta}_{{\text{G}},i})\sum\nolimits_{k=1}^4 {\ket{k}_s\otimes\ket{k}_i}.
\end{equation}

\subsection{Discriminator's circuit: performing arbitrary projections on each ququart}

\begin{figure*}[!h]
  \centering\includegraphics{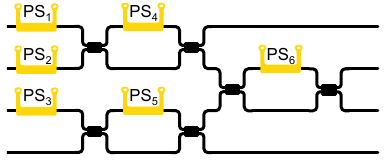}
  \caption{\centering Schematic of the 4-mode projection circuit.}
  \label{fig: P}
\end{figure*}

In Ref. \cite{Wang2023}, we have proposed a scheme to perform arbitrary projective measurement upon the on-chip single-ququart states. At first, we give a brief review. Any single ququart state can be decomposed into the parametric form: 
\begin{equation} \label{Eq: arbitrary ququart form}
  \begin{split}
    \ket{\psi(\vec{\theta})}
    &=\ket{\psi(\theta_{1},\theta_{2},\theta_{3},\phi_{1},\phi_{2},\phi_{3},\phi_{4})} \\
    &= e^{i\phi_{1}} \sin\theta_{1}\sin\theta_{2} \ket{1} + 
    e^{i\phi_{2}} \sin\theta_{1}\cos\theta_{2} \ket{2} 
    + e^{i\phi_{3}} \cos\theta_{1}\sin\theta_{3} \ket{3} + 
    e^{i\phi_{4}} \cos\theta_{1}\cos\theta_{3} \ket{4},
  \end{split}
\end{equation}
where $\ket{k}$ ($k$=1,2,3,4) are the path modes of the single photons.
Given a projector $\hat{P}(\vec{\theta}_\text{M})=\ket{\psi_\text{M}(\vec{\theta}_\text{M})}\bra{\psi_\text{M}(\vec{\theta}_\text{M})}$ described by an arbitrary single-ququart basis $\ket{\psi_M(\vec{\theta}_\text{M})}$, the measurement upon the single ququart state $\ket{\psi_\text{S}(\vec{\theta}_\text{S})}$ evolved after the preceded optical circuit can be performed by the structure in Fig. \ref{fig: P}. With the six phase shifters set as
\begin{equation} \label{Eq: phase setting}
  \begin{tabular}{ll}
    $\varphi_{{\rm PS}_{1}}=-\phi_\text{M1}+\phi_\text{M4}-\theta_\text{M2}+\theta_\text{M3}+\frac{\pi}{2},$ & $\varphi_{{\rm PS}_{4}}=\pi+2\theta_\text{M2},$ \\
    $\varphi_{{\rm PS}_{2}}=-\phi_\text{M2}+\phi_\text{M4}-\theta_\text{M2}+\theta_\text{M3}+\frac{\pi}{2},$ &  $\varphi_{{\rm PS}_{5}}=2\theta_\text{M3},$ \\
    $\varphi_{{\rm PS}_{3}}=-\phi_\text{M3}+\phi_\text{M4},$ & $\varphi_{{\rm PS}_{6}}=2\theta_\text{M1},$\\
  \end{tabular}
\end{equation}
the operation $\hat{U}_\text{proj}=\hat{U}_\text{proj}(\varphi_{{\rm PS}_{1}},\varphi_{{\rm PS}_{2}},\varphi_{{\rm PS}_{3}},\varphi_{{\rm PS}_{4}},\varphi_{{\rm PS}_{5}},\varphi_{{\rm PS}_{6}})$ has a relation $U_\text{proj}\ket{\psi_\text{M}(\vec{\theta}_\text{M})}=e^{i(\theta_\text{M1}+\theta_\text{M3}+\phi_\text{M4}+\pi)} \ket{2}$, then the detection probability of the photon at the second output port is:
\begin{equation}
  \begin{split}
    \text{P}_{(2)} =& \left|{ \bra{2}U_\text{proj}\ket{\psi_{\text{prep}}} }\right|^2 \\
    =&\left|{\bra{\psi_\text{M}(\vec{\theta}_\text{M})}{U_\text{proj}}^\dagger U_\text{proj}\ket{\psi_\text{S}(\vec{\theta}_\text{S})}}\right|^2\\
    =&\left|{\braket{\psi_\text{M}(\vec{\theta}_\text{M})|\psi_\text{S}(\vec{\theta}_\text{S})}}\right|^2\\
    =&{\bra{\psi_\text{S}(\vec{\theta}_\text{S})}\hat{P}(\vec{\theta}_\text{M})\ket{\psi_\text{S}(\vec{\theta}_\text{S})}}.
  \end{split}
\end{equation}
Accordingly, the average value of the measurement of the projector $\hat{P}(\vec{\theta}_\text{M})$ upon the on-chip single-ququart states can be experimentally estimated.

Next, we generalize this scheme to the entangled two-ququart state $\ket{g(\vec{\theta}_{\text{G}})}$ from the generator's circuit. A projective measurement on $\ket{g(\vec{\theta}_{\text{G}})}$ can be described by the two-ququart projector
\begin{equation}
  \mathcal{\hat{M}}(\vec{\theta}_{\text{D}})=\hat{P}_s(\vec{\theta}_{{\text{D}},s}) \otimes \hat{P}_i(\vec{\theta}_{{\text{D}},i}),
\end{equation}
where $\hat{P}_s(\vec{\theta}_{{\text{D}},s})=\ket{\psi(\vec{\theta}_{{\text{D}},s})}\bra{\psi(\vec{\theta}_{{\text{D}},s})}$ and $\hat{P}_i(\vec{\theta}_{{\text{D}},i})=\ket{\psi(\vec{\theta}_{{\text{D}},i})}\bra{\psi(\vec{\theta}_{{\text{D}},i})}$ are respective projectors of the signal and idler photons. According to Eq. (\ref{Eq: phase setting}), the six phase shifters in the signal photon measurement circuit and the six phase shifters in the idler photon measurement circuit can be configured. Then, the transformations on the signal and idler photons, $\hat{U}_{\text{proj},s}$ and $\hat{U}_{\text{proj},i}$, have the relations $U_{\text{proj},s}\ket{\psi(\vec{\theta}_{\text{D},s})}=e^{i\theta_s} \ket{2}_s$ and $U_{\text{proj},i}\ket{\psi(\vec{\theta}_{\text{D},i})}=e^{i\theta_i} \ket{2}_i$, where $\theta_s$ and $\theta_i$ are some global phases. Then, the probability that the signal and idler photons are coincidentally detected at both second ports is
\begin{equation}\label{Eq: two-ququart projector}
  \begin{split}
    \text{P}_{(2,2)} =& \left|{ \left(\bra{2}_s \otimes \bra{2}_i\right) \hat{U}_{\text{proj},s} \otimes \hat{U}_{\text{proj},i}  \ket{g(\vec{\theta}_{\text{G}})}}\right|^2 \\
    =&\left|{\left(\bra{\psi(\vec{\theta}_{{\text{D}},s})}\otimes\bra{\psi(\vec{\theta}_{{\text{D}},i})}\right) \left({U_{\text{proj},s}}^\dagger \otimes {U_{\text{proj},i}}^\dagger\right) \left(\hat{U}_{\text{proj},s} \otimes \hat{U}_{\text{proj},i}\right)  \ket{g(\vec{\theta}_{\text{G}})}}\right|^2 \\
    =&\left|{\left(\bra{\psi(\vec{\theta}_{{\text{D}},s})}\otimes\bra{\psi(\vec{\theta}_{{\text{D}},i})}\right) \ket{g(\vec{\theta}_{\text{G}})}}\right|^2\\
    =&{\bra{g(\vec{\theta}_{\text{G}})} \hat{P}_s(\vec{\theta}_{{\text{D}},s}) \otimes \hat{P}_i(\vec{\theta}_{{\text{D}},i}) \ket{g(\vec{\theta}_{\text{G}})}}\\
    =&{\bra{g(\vec{\theta}_{\text{G}})} \mathcal{\hat{M}}(\vec{\theta}_{\text{D}}) \ket{g(\vec{\theta}_{\text{G}})}}.
  \end{split}
\end{equation}
Thus, the measurement result of the two-ququart projector $\mathcal{\hat{M}}(\vec{\theta}_{\text{D}})$ on $\ket{g(\vec{\theta}_{\text{G}})}$ can be estimated experimentally by using the projection circuits.

In the quantum generative adversarial learning protocol, the discriminator does not need to know the fidelity of the generated state, let alone the complete state information obtained by quantum tomography. As a witness of the learning process, we use these circuits to estimate the quantum state fidelity of the experimentally obtained state $\hat{\rho}(\vec{\theta}_{\text{G}})=\ket{g(\vec{\theta}_{\text{G}})}\bra{g(\vec{\theta}_{\text{G}})}$ and theoretical state $\hat{\tau}$ \cite{Nielsen2010}
\begin{equation}\label{Eq: two-photon coincidence}
  \begin{split}
    \text{F}_q[\hat{\rho}(\vec{\theta}_{\text{G}}),\hat{\tau}] 
    &= \text{tr}[\sqrt{\hat{\rho}^{1/2}(\vec{\theta}_{\text{G}})\hat{\tau}\hat{\rho}^{1/2}(\vec{\theta}_{\text{G}})}] \\
    &=\sqrt{\bra{g(\vec{\theta}_{\text{G}})}\hat{\tau}\ket{g(\vec{\theta}_{\text{G}})}}.
  \end{split}
\end{equation}
Specifically, the arbitrary density operator $\hat{\tau}$ of a two-ququart state is a Hermitian operator and has a spectral decomposition $\hat{\tau} = \sum_{j=1}^{16^2} {m_j \mathcal{\hat{M}}_j}$,
where $\mathcal{\hat{M}}_j$ are two-ququart projectors. Following Eqs. (\ref{Eq: two-ququart projector}) and (\ref{Eq: two-photon coincidence}), we can measure the average value of ${\bra{g(\vec{\theta}_{\text{G}})} \mathcal{\hat{M}}_j \ket{g(\vec{\theta}_{\text{G}})}}$ by setting the two projection circuits and measuring the probability that the signal and idler photons are coincidentally detected at both second ports. Finally, the quantum state fidelity can be estimated
\begin{equation}
  \text{F}_q[\hat{\rho}(\vec{\theta}_{\text{G}}),\hat{\tau}] = \sqrt{\sum_{j=1}^{16} {m_j {\bra{g(\vec{\theta}_{\text{G}})} \mathcal{\hat{M}}_j \ket{g(\vec{\theta}_{\text{G}})}}}}.
\end{equation}

\section{Device fabrication}

Our silicon chip is designed and fabricated on a silicon-on-insulator (SOI) platform with a top crystalline silicon thickness of 220 nm and a buried oxide (BOX) layer of 2 $\mu$m. The 500-nm-wide silicon waveguides were patterned in the 220-nm top layer of silicon. Resistive heaters that are 180 $\mu$m long and 2.5 $\mu$m wide and work as thermo-optic tunable phase shifters are then patterned on a 120 nm TiN metal layer on top of the waveguide layer. 
Finally, the fabricated and cleaved chip was wire-bonded to a printed circuit board.

\section{Experimental setup}

\begin{figure*}[!h]
  \centering\includegraphics[width=14cm]{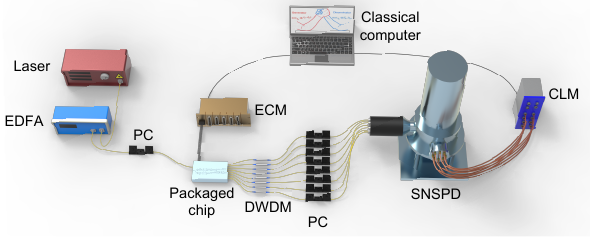}
  \caption{\centering Schematic of the experimental setup.}
  \label{fig: setup}
\end{figure*}

The schematic diagram of the whole experimental setup is shown in Fig. \ref{fig: setup}. 
A tunable continuous wave laser is tuned to a wavelength of 1549.32 nm. 
Then, the laser is amplified with an optical erbium-doped fiber amplifier (EDFA). The amplified spontaneous emission noise is suppressed using a dense wavelength-division multiplexing filter (DWDM).
The pump laser is launched into the chip (circled by yellow) through a V-groove fiber array (VGA), which is optically aligned and then firmly glued with the chip using the optical adhesive. 
A polarization controller (PC) ensures that TE-polarized light is launched into the chip through on-chip TE grating couplers.
Photon pairs generated from the chip are collected from the same VGA. 
The non-degenerate signal (1554.13 nm, denoted in blue) and idler (1544.53 nm, denoted in red) photons are separated by off-chip DWDM modules. Then, photons are detected by superconducting nanowire single-photon detectors (SNSPDs). The polarization of photons before going into SNSPDs is optimized by PCs, which maximizes the detection efficiency up to around $85\%$. 
Two-photon coincidence events are recorded by a coincidence logic module (CLM) with a resolution of 156.25 ps time bin width. During our experiments, the dark count for single photon detection is around 300 per second and is ignorable for two-photon coincidence events. 
An electrical control module (ECM) configures all the on-chip thermo-optic phase shifters via electrical pads placed on three sides of the chip that are wire-bonded to a multi-layered printed circuit board (PCB), with vulnerable gold wires protected by solid insulation adhesive. 
A classical computer compiles the QGAN protocol and coordinates control and measurement modules.

\section{Device performance}
\subsection{Device Loss}
The overall insertion loss of the chip is estimated as 20 dB for the light passing through the sources to one output of the grating coupler. We estimate losses based on measurements on the test structures on the same die. The propagation loss for the waveguide is 3 dB/cm. The waveguide crosser has around 0.1 dB insertion loss. The multi-mode interferometer (MMI) has around 0.3 dB insertion loss. The thermo-optic phase shifters have ignorable loss compared to the MMIs. 
The fabricated grating coupler reaches the highest coupling efficiency of around -4 dB with a light of 1549.32 nm.

\subsection{Performance of photon sources}
\begin{figure*}[!h]
  \centering\includegraphics{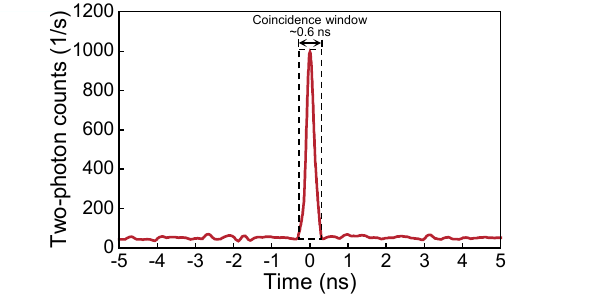}
  \caption{\centering Histogram of the two-photon coincidence counts.}
  \label{fig: coincidence}
\end{figure*}

Upon using a 50 mW pump laser, each source can generate approximately 1500 two-photon coincidence counts per second, within a coincidence window of 0.625 ns. Fig. \ref{fig: coincidence} is a typical histogram of coincidence at various time delays. The multiphoton emission rate can be ignorable in our experiments.

\begin{figure*}[!h]
  \centering\includegraphics[width=18cm]{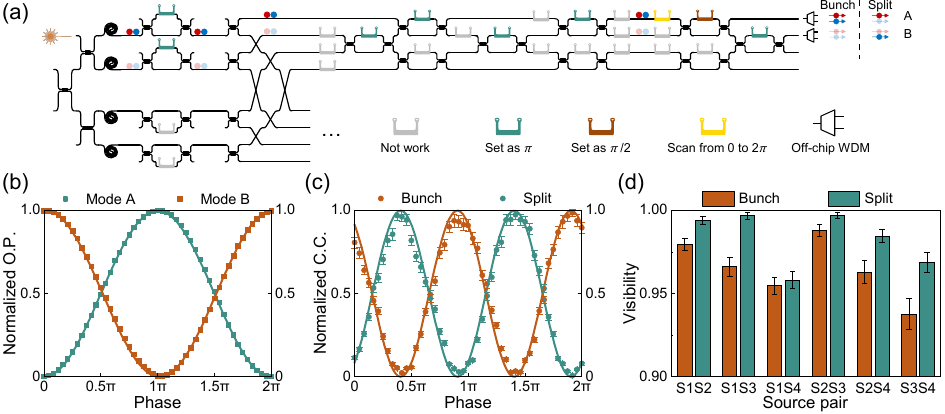}
  \caption{On-chip quantum and classical interference measurements.(a) The configuration for the reverse Hong-Ou-Mandel (RHOM) fringes on our chip. The separation of signal (blue) and idler (red) photons is achieved using off-chip WDMs. Here, the bunch state describes photons bunched together in either output mode A or B, while the split state refers to a pair of photons where one is present in each mode. (b) Transmission of the bright laser shows classical interference ($V=99.87\%$ for the output from mode A and $V=99.42\%$ for the output from mode B).  (c) Measurement of signal-idler photon bunching with signal and idler both in modes A or B ($V=97.96\pm0.37\%$) and splitting between modes A and B ($V=99.43\pm0.24\%$), showing quantum interference. The points are experimental data, while the lines are obtained by sinusoidal and cosinoidal fittings. Error bars are given by Poissonian statistics of photons.(d) Visibilities of different choices of pairs of sources are all above $90\%$, showing a good level of indistinguishability even after the deep circuits. }
  \label{fig: HOM}
\end{figure*}

Sources distinguishability would insert other noise in the generated states and therefore affecting computing performances. The effects of source noise on the QGAN performance warrant further systematic study. Here, to quantify the indistinguishability of the four non-degenerate sources after deep circuits, we performed two-photon reverse Hong-Ou-Mandel (RHOM) interference \cite{Silverstone2014,Wang2018,Vigliar2021,Bao2023} experiments at the end of our circuits, for all possible six pairs of the four sources. 
The configuration for the chip to quantify the indistinguishability between the first and second sources is shown in Fig. \ref{fig: HOM}.
This interference is actually the time reversal of the Hong-Ou-Mandel experiment: two non-degenerate photons that come from the same source and are in the same spatial mode impinge at a beam splitter and come out at different ports, resulting in a coincidence peak \cite{Vigliar2021}. 
By scanning the phase between the path modes of the two sources and performing two-photon coincidence detection on the two outputs, we obtain a quantum interference fringe. 
When compared to their classical counterparts in Fig. \ref{fig: HOM}(b), the quantum fringes in Fig. \ref{fig: HOM}(c) have the significant phase doubling. 
The quantum interference exhibited a high visibility of $V=97.96\pm0.37\%$ for the splitting fringe and $V=99.43\pm0.24\%$ for the bunching fringe. Here, visibility is defined as $V=\frac{N_\text{max}-N_\text{min}}{N_\text{max}}$ from the maximum count $N_\text{max}$ and the minimum count $N_\text{min}$ \cite{Silverstone2014}. By reconfiguring the chip, we performed RHOM interference between different pairs of sources and obtained a pairwise characterization of their  indistinguishability. As shown in Fig. \ref{fig: HOM}(d), the measured visibilities of the RHOM fringes are all above $90\%$, showing a high level of indistinguishability.

\subsection{Phase-shifter calibration}

We use a self-developed multi-channel electrical control module (ECM) to drive the on-chip thermal-optic phase shifters. The ECM can respond to the device driver and send analog electrical signals to reconfigure all the on-chip thermo-optic phase shifters parallelly.
The resistance value of each on-chip phase shifter can be measured by scanning its I-V curve. The fitted resistances of phase shifters in the universal linear optical circuits and measurement circuits on our silicon chip are between $770\sim820\Omega$. 

Each thermal-optic phase shifter has a linear phase electrical-power relationship and further a nonlinear phase voltage relationship, considering the fact that electrical-power $P$ of a phase shifter is given by
\begin{equation}
  P=V^2/R,
\end{equation}
where $V$ represents the applied voltage and $R$ represents the resistance of the phase shifter. Thus, a nonlinear phase voltage relationship is given as \cite{Qiang2018,Qiang2021}
\begin{equation} \label{Eq: nonlinear phase voltage relationship}
  \theta(V)=\phi_0 + \phi_1 V^2,
\end{equation}
where $\theta(V)$ is the resulting phase shift. $\phi_0$ and $\phi_1$ are real numbers associated with the response of a particular heater.

\begin{figure*}[!t]
  \centering\includegraphics{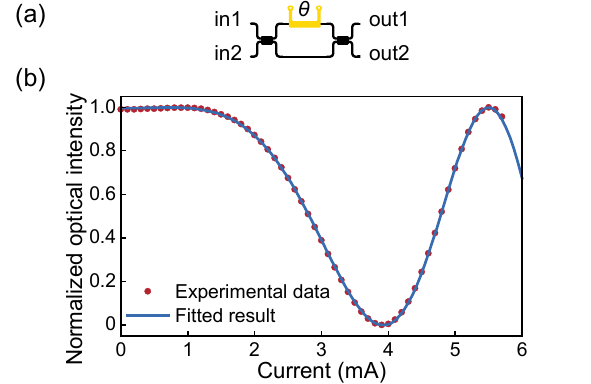}
  \caption{(a) Schematic diagram of a phase shifter. (b) The voltage optical-intensity fringe curve of an example phase shifter. The red points represent the experimental data and the blue curve is obtained through fitting Eq. (\ref{Eq: nonlinear phase voltage relationship}), showing good fitness.}
  \label{fig: cali}
\end{figure*}

To calibrate the phase shifter in the circuits, we inject a bright laser (1550 nm, about 1 mW) into one input port (In1) of the MZI and measure the intensity at one output port (Out2) as a function of the voltage applied to the heater. The applied voltage increases linearly from 0V to 5.7V (0.1V stepsize). The produced classical interference fringe of one example phase shifter is shown in Fig. \ref{fig: cali}. Using nonlinear fitting, we are able to obtain the fitting parameters $\phi_0$ and $\phi_1$ for each heater. For the specific case shown in Fig. \ref{fig: cali}, we obtained that $\phi_0 = 0.2112$, $\phi_1 = 6.1728$.

All the thermo-optic phase shifters in the universal linear optical circuits and measurement circuits each have about $800\Omega$ resistance and the applied current for $2\pi$ phase shift is less than 6 mA. The electric power required for $2\pi$ phase shift for one individual thermo-optic phase shifter is around 50 mW. According to the calibration result in Fig. \ref{fig: cali}, the extinction ratio of the MZI where the calibrated phase shifter is located within was obtained to be around 40 dB, which is the typical value of the extinction ratio on our chip.

\subsection{Performance of universal linear optical circuits}

\begin{figure*}[!h]
  \centering\includegraphics{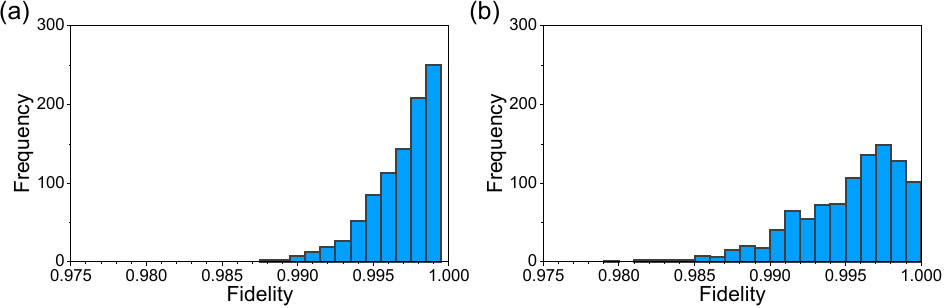}
  \caption{Histograms of the measured classical fidelities for 1000 Haar-randomly generated 4-dimensional unitary operations implemented in universal linear optical circuits of (a) signal photon and (b) idler photon.}
  \label{fig: Classical fidelity}
\end{figure*}

An arbitrary 4-mode unitary transformation can be implemented with the universal linear optical circuits as described in Fig. \ref{fig: U}. Two copies of the same universal linear optical circuits are fabricated on our chip. To showcase their performance, we programmed each universal linear optical circuit to implement 1000 Haar-randomly generated 4-dimensional unitary operations with the input state as $[0.5,0.5,0.5,0.5]^\text{T}$.
We measure the output optical intensity distribution for each circuit. The classical fidelity $F_c$ between the experimentally obtained distribution and the theoretical result is defined as
\begin{equation}
    {F_c}[{\vec{P}_{\rm{th}}},{\vec{P}_{{\rm{exp}}}}] = \sum\limits_{i = 1}^{4} {\sqrt {{{P}_{\rm{th},i}}{{P}_{{\rm{exp},i}}}}},
\end{equation}
where $\vec{P}_{\rm{th}}$ and $\vec{P}_{\rm{exp}}$ represent the theoretical and experimental output optical distribution vector, respectively. The statistical fidelities for the signal photon circuit and the idler photon circuit are as high as $99.71\pm0.22\%$, $99.51\pm0.36\%$, respectively. The histograms of measured fidelities for each circuit are shown in Fig. \ref{fig: Classical fidelity}.

\section{Noise in control and detection of photonic chips}

\subsection{Phase noise}
On-chip phase shifters are essential components in quantum photonic circuits, enabling precise control and manipulation of the on-chip phase of photons \cite{Wang2020a}. However, these phase shifters are susceptible to various noise sources that can degrade their performance and impact the overall quantum operations of the chip. We will discuss some crucial noise sources that affect on-chip phase shifters. 

\subsubsection{Thermal crosstalk}
Our on-chip phase shifters are implemented using thermo-optic effects, where the refractive index of the waveguide is modulated by temperature. However, the heat crosstalk between adjacent thermo-optical phase shifters and thermal fluctuations inherently affects these modulation mechanisms, leading to phase noise. Thermal noise arises due to random thermal energy fluctuations in the material, causing variations in the refractive index and subsequently introducing phase errors. In our chip, the distance between parallel phase shifters is 120 $\mu$m, and the distance between cascaded phase shifters is 340 $\mu$m. In the calibration process of phase shifters, we confirmed that the crosstalk between neighboring phase shifters on our chip is ignorable, by applying voltage to a neighboring phase shifter. There is no doubt that smaller gaps between phase shifters will introduce more and stronger unwanted crosstalk. The investigation and mitigation of thermal crosstalk on photonic chips is of great significance in integrated photonics. Many designs and methods have been proposed to investigate the complex thermal crosstalk on large-scale photonic chips and design the dense photonic integration with low thermal crosstalk \cite{Gilardi2014,Milanizadeh2019,De2020}.

\subsubsection{Control deviation}
The control noise from the electrical control module has two effects. Firstly, the accuracy of phase-shifter calibration depends on the precision of the driven voltage. The voltage deviation in the calibration process will introduce perturbation in the voltage optical-intensity fringe curve, leading unwanted offsets in the nonlinear phase voltage relationship shown in Eq. (\ref{Eq: nonlinear phase voltage relationship}). This has a lasting offset impact on the phase shifter.
Secondly, when driving more phase shifters, the power supply of the electrical control module can show more fluctuations. This introduces inaccuracies to the configured target voltage of each phase shifter and further reduces the fidelities of experimental obtained operations.

\subsubsection{Environmental noise}
On-chip phase shifters are operated in real-world environments, which can introduce various external noise sources. Environmental factors such as temperature variations and vibrations can impact the performance of the phase shifters. These noise sources can couple with the device, leading to phase errors.

\subsubsection{Characterizing phase noises}
Understanding and characterizing the sources of phase noise is crucial for designing and optimizing on-chip phase shifters in integrated photonic quantum chips. However, analytically characterizing the noises on a realistic photonic chip can be challenging due to \textit{complex noise interactions}. On a photonic chip, multiple noise sources, including the above three and even other unknown sources, can interact, making it difficult to isolate and analyze individual noise contributions. The interplay between different noise sources can lead to complex noise correlations and dependencies, making analytical characterization challenging. 

One thing, however, is clear: all these different noise sources will introduce phase noise to the target phase $\theta$, and the stronger the noise sources, the greater the phase noise level. Therefore, following Ref. \cite{Paesani2017a}, we use the Gaussian noise model with $\theta^{\prime} \sim \mathcal{N}(\theta,\sigma^2)$ to describe the actual phase $\theta^{\prime}$ affected by different noise level $\sigma$. This allows us to succinctly and effectively examine the robustness of QGAN to different phase noise intensities, which is the focus of our work, without having to characterize the complex effect of each noise on the phase shifter.

\subsection{Shot noise}
In silicon quantum photonics, photon sources are based on spontaneous four-wave mixing (SFWM) \cite{Silverstone2014}. This probabilistic scheme produces entangled multiple photons at an exponentially decaying rate with the photon number \cite{Adcock2021}. In some state-of-the-art chips \cite{Vigliar2021,Bao2023}, the four-photon rate  is as low as $0.001\sim1$ Hz. Besides, the insertion loss of on-chip building blocks such as waveguide and MMI, coupling loss of couplers, and detection loss of detectors can further reduce the photon rate.
Thus, when the QGAN on a silicon quantum photonic chip scales up to more photons and larger sizes, it is inevitable to face a low data rate. 

When detecting photons at a low rate, shot noise arises from the discrete nature of photons \cite{Campbell1909a,Campbell1909b,Eliazar2005}. When a multiphoton event is detected, there is inherent uncertainty in the exact time of its arrival. The number of arriving multiphoton per unit of time is stochastic, leading to random fluctuations in the number of multiphoton detected.
If $N$ is the number of the multiphoton detected, the Poisson distribution gives the fluctuations by the standard deviation, that is equal to the mean multiphoton count
\begin{equation}
  \sigma = \sqrt{N}.
\end{equation}
The average count divided by the standard deviation of the fluctuations is the signal to noise ratio (SNR)
\begin{equation}
  \text{SNR} = \frac{N}{\sigma} = \sqrt{N}.
\end{equation}
Therefore, when the integration time is limited, the low multiphoton count and the intense shot noise degrade the SNR and the precision of the obtained  probability distribution.

\begin{figure*}[!h]
  \centering\includegraphics[width=18cm]{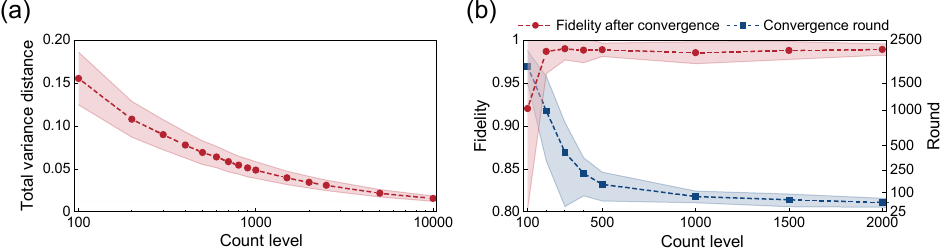}
  \caption{Effect of shot noise in photonics on (a) the obtained probability distribution and (b) the performance of  quantum generative adversarial learning.}
  \label{fig: Shot noise}
\end{figure*}

Here, we give some examples demonstrating the effects of shot noise on the precision of measured probability distribution. Given a theoretical probability distribution $\vec{P}_\text{the}$, which is a 16-dimensional uniform distribution. We numerically simulated the experimentally obtained distribution $\vec{P}_\text{exp}$ affected by shot noise with a total count ranging from 100 to 10000. We calculate the total variance distance between $\vec{P}_\text{the}$ and $\vec{P}_\text{exp}$
\begin{equation}
  \text{TVD}=\left\| {\vec{P}_\text{the} -\vec{P}_\text{exp}} \right\| = \frac{1}{2}\sum\nolimits_{i=1}^{16} {\left| {{P_\text{the,i}} - {P_\text{exp,i}}} \right|}.
\end{equation}
For each count level, we repeat the simulation 1000 times and obtain the average of the TVD. The results in Fig. \ref{fig: Shot noise}(a) clearly show that low count levels degrade the precision of the obtained probability distribution. In our experiments, we set the two-photon integration time as one second to ensure a total count of around 1500.

For quantum generative learning in photonics, shot noise increases the instability of the learning process. We demonstrate the effects of shot noise on the learning result by performing 100 numerical simulations of quantum generative adversarial learning in photonics with a total count ranging from 100 to 10000. The phase noise is set as 0. The results are presented in Fig. \ref{fig: Shot noise}(b). It is obvious that low count level and, thus, intense shot noise can degrade the performance of QGAN, leading to the reduction in the quality of the generated quantum state and the increase in the learning round.

\section{Effect of circuit defects on Quantum generative adversarial learning}
According to Schmidt decomposition \cite{Nielsen2010}, preparing a two-body maximally entangled state requires applying the corresponding unitary transformation on the maximally entangled state. Currently, the generator circuit consists of two universal linear optical networks capable of performing arbitrary unitary transformations, and each phase shifter is indispensable in its respective linear optical network. In quantum generative adversarial learning, if the task of the generator is to prepare an arbitrary unknown two-body maximally entangled state, then none of the phase shifters can be defective; otherwise, it may not be possible to accurately perform the corresponding unitary transformation.

For quantum generative adversarial learning in photonics, an interesting question arises: whether there exists a crucial phase shifter that, once broken, would significantly affect the quality of generated quantum states. If the answer is "yes," we can remove the non-essential phase shifters and achieve more compact circuits.  We search for such crucial phase shifters through extensive numerical simulations. For example, when testing the learning performance with one defective phase shifter, we randomly select one phase shifter as the defective phase shifter from the 15 phase shifters in each universal linear optical circuit. From 100 simulation results, it can be observed that compared to the final fidelity without defects ($99.14\pm1.23\%$), the average fidelity of the generated quantum states slightly decreases ($98.33\pm7.00\%$), but no single phase shifter will have a significant impact on fidelity if it is defective.
This means that after one phase shifter is damaged, other phase shifters can still adapt to this disadvantageous situation by cooperating with others, resulting in a not bad learning performance. This is an intriguing phenomenon of unity and mutual assistance. Mathematically, this corresponds to an interesting problem: how to perform unitary decomposition under restricted conditions. 
However, the fidelity gradually decreases as the number of defects (randomly selected) increases, indicating that the adaptability is limited. This implies that the more constraints there are, the more complex the unitary decomposition becomes and the worse the learning performance.
Accurate preparation of entangled quantum states with fewer hardware resources is an essential problem for quantum computing and quantum information processing. It is worth continuing our research to determine whether a minimal structure is capable of accomplishing this task.

\clearpage

\bibliography{supplement}